\documentclass[prl,twocolumn,epsf]{revtex4}
\usepackage{graphicx}% Include figure files
\usepackage{dcolumn}% Align table columns on decimal point
\usepackage{bm}% bold math
\usepackage{hyperref}% Auto-generate hyperlinks

\newcommand{\beq}{\begin{eqnarray}}
\newcommand{\eeq}{\end{eqnarray}}
\begin{document}
\noindent
{\bf Comment on ``High-Spin Polaron in Lightly Doped CuO$_2$ Planes'' }

In a recent Letter,  Lau, et al.\cite{lau} investigated the single hole problem in an effective model containing Cu $d$ and O $2p$ orbitals by 
exact diagonalization on clusters composed of 32 CuO$_2$ unit cells. With full quantum fluctuations due to the antiferromagnetic (AFM)
backgroud, they found that spin-polaron solutions with spin 3/2 become lowest energy solutions in certain region of Brillouin zone (BZ), 
which they claimed no similiar solutions were obtained in other models or approximations before. 
In this Comment, we would like to point out that such high-spin polaron solutions 
have been found in our previous Letter\cite{leewc2003} published in 2003 and many of features described by Lau et al. have in fact 
been seen in our Letter.

Starting from the single-band $t-J$ model, we constructed the variational wave function (VWF) for low energy states with single-hole in the AFM
background as:
\begin{eqnarray}
|\Psi_1({\bf q}_{s})\rangle & = &
P_d~c^{\dagger}_{{\bf q}_{s}\uparrow} \nonumber \\
& & \mbox{} [{\sum_{[{\bf k} \neq {\bf q}_{h}]} 
(A_{\bf k} a^{\dagger}_{{\bf k}\uparrow}a^{\dagger}_{{\bf -k}\downarrow}
 +B_{\bf k} b^{\dagger}_{{\bf k}\uparrow}b^{\dagger}_{{\bf -k}\downarrow})}
]^{(N_s/2)-1}
|0\rangle ,\nonumber\\
\label{twf-sb}
\end{eqnarray}
Detailed description of the symbols used in Eq. \ref{twf-sb} can be found in our Letter\cite{leewc2003}.
${\bf q}_{h}$ denotes the hole momentum which is excluded from the sum
if ${\bf q}_{h}$ is within the spin BZ,
otherwise, ${\bf q}_{h}-{\bf Q}$ is excluded where ${\bf Q}=(\pi,\pi)$ is the AFM wave vector. 
${\bf q}_{s}$ is the momentum of unpaired spin which is present due to the creation of a single hole.
$P_d$ here enforces the constraint of no doubly occupied sites. 

As shown in Refs. [\onlinecite{lee-shih, leewc2003}], the quasi-particle (QP) states with finite spectral weight can be obtained from Eq. \ref{twf-sb} by setting ${\bf q}_{h}={\bf q}_{s}$. Energy dispersion of these QP states is quite consistent with the exact solutions of 32 sites for the $t-J$ model\cite{leung}
and extended $t-J$ model\cite{lee-shih,leung2}. Comparing the energy dispersion and the spectral weight presented in 
Figs. 2 and 4 in Ref. [\onlinecite{lee-shih}] with those shown in Fig. 2 in Lau's Letter\cite{lau}, it is clear that 
these two results also agree with each other nicely. Especially, the features that QP states have the lowest energy at $(\frac{\pi}{2},\frac{\pi}{2})$
and smallest spectral weight at $(\pi,\pi)$ are well-captured by our VWFs.

In addition to QP states, we obtained another different set of states, namely the {\it spin-bag} (SB) states, by choosing ${\bf q}_{h}\neq{\bf q}_{s}$.
As explained in our Letter\cite{leewc2003}, the SB states can be viewed as exciting a spin wave with momentum ${\bf k'}={\bf q}_{s} - {\bf q}_{h}$ on top 
of a QP state with momentum ${\bf q}_{h}$. These SB states carrying almost negligible spectral weight was also discussed in the paper. This construction is in complete agreement with the concept of high-spin polaron 3/2 states presented in Lau's Letter. 

Furthermore, we found that the spin-bag states could be the lowest energy states in certain region of BZ, and interestingly the discussion given in Lau's Letter 
about the energy difference between spin 3/2 and 1/2 states along the direction from $(0,0)$ to $(\frac{\pi}{2},\frac{\pi}{2})$ and then to $(\pi,\pi)$ 
matches almost perfectly with our results on QP and SB states shown in Fig. 1 in our Letter\cite{leewc2003}.

By assuming that a Zhang-Rice singlet\cite{zr1988} could be formed by superimposing the weight of four nearest neighbor oxygen, we could calculate the spin-spin correlations based on Lau's Fig. 3(a). The values of -0.225 and -0.273 are obtained for the e and d bonds defined in the inset of Fig.2 in \cite{leewc2003}. These are fairly compatible with our data of -0.215 and -0.224 respectively, although different parameters were used in the two calculations. 

Since almost all the low energy states presented by Lau et al. are consistent with our results on a singe-band $t-J$ model, 
the claim of the breakdown of Zhang-Rice singlet is not convincing. In our viewpoints, the smoking gun signature proving the breakdown of Zhang-Rice singlet 
should be a calculation showing that non-bonding state of Cu spin with its four neighboring oxygen holes indeed has lower or comparable energy with the Zhang-Rice singlet. 
Such a calculation, up to our knowledge, has never been reported to date.

{\it Acknowledgement} -- W.-C. Lee would like to thank Aspen Center for Physics supported by NSF grant No. 1066293, USA and T.K. Lee would like to thank 
Max Planck Institute for the Physics of Complex Systems for their hosting when this Comment is finalized. 

\vspace{12pt}
\noindent Wei-Cheng Lee,
Department of Physics\\
University of Illinois,
Urbana, IL, 61801-3080\\
T. K. Lee,
Institute of Physics,\\
Academia Sinica,
Nangang, Taipei, Taiwan

\end{document}